\documentstyle[12pt]{article}
\begin{document}
\title{NO TIME ASYMMETRY FROM QUANTUM
    MECHANICS\thanks{Alberta-Thy-11-93, gr-qc/9304027}}
\author{ Don N. Page\\
CIAR Cosmology Program\\
Theoretical Physics Institute\\
Department of Physics\\University of Alberta\\
Edmonton, Alberta\\Canada T6G 2J1\\
Internet:  don@page.phys.ualberta.ca}
\date{(1993 Feb. 12, revised April 19)}
\maketitle
\large
\begin{abstract}
\baselineskip 25pt
With CPT-invariant initial conditions that commute with
CPT-invariant final conditions, the respective probabilities
(when defined) of a set of
histories and its CPT reverse are equal,
giving a CPT-symmetric universe.  This leads me to
question whether the asymmetry of the
Gell-Mann--Hartle decoherence functional
for ordinary quantum mechanics should be
interpreted as an asymmetry of {\it time} .
\end{abstract}
\normalsize
\pagebreak
\baselineskip 24.2pt
There are many time asymmetries observed in our universe
(not all unrelated),
such as the thermodynamic arrow of time, the arrow of
retarded radiation, the psychological arrow, the expansion of
the universe, and the T-noninvariance of the \( K^{0}\)
system.  The collapse of the wavefunction through the
process of measurement \cite{1} has sometimes appeared to
be an independent quantum arrow of time \cite{2}, though it
has also been ascribed to the thermodynamic time
asymmetry of the external measuring apparatus or
environment \cite{3}.

Aharonov, Bergmann, and Lebowitz \cite{4} have proposed a
time-symmetric generalization of ordinary quantum
mechanics by using ensembles of histories with both initial
and final states.  Griffiths \cite{5}, and later Unruh \cite{u}
and then Gell-Mann and Hartle \cite{6,7,l}, have developed a similar
formulation in terms of an initial and a final density matrix. In
this
formulation, ordinary quantum mechanics corresponds to the
case in which the final density matrix is proportional to the
identity, which denotes a final condition of indifference and
which Gell-Mann and Hartle argue gives ordinary quantum
mechanics an arrow of time.

Here I shall prove a theorem implying the CPT-invariance of
probabilities in ordinary quantum mechanics
when the initial density matrix is CPT invariant,
which is thus sufficient to give a CPT-invariant universe,
assuming, as I shall do throughout, that the Hamiltonian
is CPT invariant.
I shall follow this with some speculative interpretations
of the asymmetry of the Gell-Mann--Hartle formulation
of ordinary quantum mechanics.

Gell-Mann and Hartle \cite{8,9} formulate the laws of
generalized quantum mechanics for a closed system in
terms of a decoherence functional
	\begin{equation}
	D(\alpha,\alpha')
	=Tr(\rho_{f}C_{\alpha}\rho_{i}C^{\dagger}_{\alpha'})
	/Tr(\rho_{f}\rho_{i}),
	\end{equation}
where $\rho_{i}$ is an initial density matrix, $\rho_{f}$ is a
final density matrix, and
	\begin{equation}
	C_{\alpha}=P^{n}_{\alpha_{n}}(t_{n})\cdots
	P^{1}_{\alpha_{1}}(t_{1})
	\end{equation}
is a string of projection operators representing the history
$\alpha=(\alpha_{1},\ldots,\alpha_{n})$ in the Heisenberg
picture, with $t_{1}<t_{2}<\cdots<t_{n}$.  Alternatively,
$C_{\alpha}$ could be a sum of such strings.  When
$\{\alpha\}$ is an exhaustive set of histories, meaning
	\begin{equation}
	\sum_{\alpha}C_{\alpha}=I,
	\end{equation}
and when this set decoheres, meaning
	\begin{equation}
	Re\,D(\alpha,\alpha')=0 \mbox{ for }\alpha\neq\alpha',
	\end{equation}
then the diagonal elements of the decoherence functional
give the probabilities for all histories of that set:
	\begin{equation}
	p(\alpha)=D(\alpha,\alpha)=Tr(\rho_{f}C_{\alpha}\rho_{i}C^
	{\dagger}_{\alpha})
	/Tr(\rho_{f}\rho_{i}).
	\end{equation}

Ordinary quantum mechanics corresponds to the special
case of this in which $\rho_{i}$ is proportional to the density
matrix of the system and $\rho_{f}$ is proportional to the
identity matrix $I$, giving a final condition of indifference.  In
this case the difference between $\rho_{i}$ and $\rho_{f}$
leads to an asymmetric decoherence functional
$D(\alpha,\alpha')$ and set of diagonal elements $p(\alpha)$,
which Gell-Mann and Hartle interpret as the
(ordinary) quantum-mechanical arrow of time.

To be specific, suppose the initial and final density matrices
are separately CPT-invariant but not the CPT reverses of
each other (so each separate state is time symmetric,
by which I shall henceforth mean CPT-invariant rather
than T-invariant in order for the dynamical laws to be
time symmetric):
	\begin{equation}
	\rho_{i}=\Theta\rho_{i}\Theta^{-1},\;\;\rho_{f}
	=\Theta\rho_{f}\Theta^{-1},
	\end{equation}
	\begin{equation}
	\rho_{f} \neq \Theta\rho_{i}\Theta^{-1},
	\end{equation}
where $\Theta$ is the antiunitary CPT operator.  Follow Gell-
Mann and Hartle \cite{7} in defining the CPT-reversed
history $\tilde{\alpha}$ represented by the string
	\begin{equation}
	\tilde{C}_{\alpha}=\tilde{P}^{1}_{\alpha_{1}}(-t_{1})
	\cdots\tilde{P}^{n}_{\alpha_{n}}(-t_{n}),
	\end{equation}
with
	\begin{equation}
	\tilde{P}^{k}_{\alpha_{k}}(-t_{k})
	=\Theta^{-1} P^{k}_{\alpha_{k}}(t_{k})\Theta,
	\end{equation}
and with the order of the projection operators reversed to put
the earlier times on the right and the later ones on the left, $-
t_{n}<\cdots<-t_{2}<-t_{1}$.  This gives
	\begin{equation}
	\tilde{C}_{\alpha}=\Theta^{-1} C^{\dagger}_{\alpha}\Theta,
	\end{equation}
which is generally true even when  $C_{\alpha}$ is a sum of
strings (2).  Then Eq. (6) implies that
	\begin{equation}
	D(\tilde{\alpha},\tilde{\alpha}')
	=Tr(\rho_{f}\tilde{C}_{\alpha}
	\rho_{i}\tilde{C}^{\dagger}_{\alpha'})/Tr(\rho_{f}\rho_{i})
	=Tr(\rho_{i}C_{\alpha'}\rho_{f}C^{\dagger}_{\alpha})
	/Tr(\rho_{i}\rho_{f}),
	\end{equation}
which would be the complex conjugate of $D(\alpha,\alpha')$
if $\rho_{f}=\rho_{i}$ [or if $\rho_{f}
=\Theta\rho_{i}\Theta^{-1}$ if Eq. (6) is not assumed]
but generally is not if the two density matrices are not so
related.  This is the Gell-Mann--Hartle asymmetry of
quantum mechanics with differing initial and final conditions.

However, if the initial and final density matrices
$\rho_{i}$ and $\rho_{f}$ commute, and if the CPT-reversed set
of histories  $\{\tilde{\alpha}\}$
obeys the decoherence condition
	\begin{equation}
	Re\,D(\tilde{\alpha},\tilde{\alpha}')
	=0 \mbox{ for }\tilde{\alpha}\neq\tilde{\alpha}'
	\end{equation}
analogous to (4), so that the diagonal elements
	\begin{equation}
	p(\tilde{\alpha})=D(\tilde{\alpha},\tilde{\alpha})=
	Tr(\rho_{f}\tilde{C}_{\alpha}
	\rho_{i}\tilde{C}^{\dagger}_{\alpha})
	/Tr(\rho_{f}\rho_{i})
	\end{equation}
obey the sum rules necessary for them
to be interpreted as probabilities,
then the separate CPT invariance of
each density matrix implies that the respective probabilities
of the corresponding sets of CPT-related histories agree, as
the following theorem shows:

\baselineskip 24pt
{\bf Theorem 1}:  If the initial density matrix $\rho_{i}$ and the
final density matrix $\rho_{f}$ commute, if they obey Eq. (6)
and hence are each separately CPT-invariant, and if the set
of histories $\{\alpha\}$ and the corresponding CPT-reversed
set $\{\tilde{\alpha}\}$ obey Eqs. (4) and (12) and hence
decohere, then the corresponding probabilities of the
respective individual histories, $p(\alpha)$ and
$p(\tilde{\alpha})$ as given by Eqs. (5) and (13), are equal.

{\bf Proof}:  Summing the decoherence condition (4) over all
$\alpha'$ different from $\alpha$ and using the
completeness relation (3) allows one to rewrite Eq. (5) as
	\begin{equation}
	p(\alpha)=Re\,Tr(\rho_{f}C_{\alpha}\rho_{i}I)
	/Tr(\rho_{f}\rho_{i})
	=Re\,Tr(\rho_{i}\rho_{f}C_{\alpha})/Tr(\rho_{f}\rho_{i}),
	\end{equation}
where the cyclic property of the trace is used here and below
to get the $C_{\alpha}$ at the right end.  Similarly, summing
Eq. (12) over all $\tilde{\alpha}$ different from
$\tilde{\alpha}'$, using (11) and the analogue of (3), and then
dropping the prime, converts Eq. (13) into
	\begin{equation}
	p(\tilde{\alpha})
	=Re\,Tr(\rho_{f}I\rho_{i}\tilde{C}^{\dagger}_{\alpha})
	/Tr(\rho_{f}\rho_{i})
	=Re\,Tr(\rho_{f}\rho_{i}\tilde{C}^{\dagger}_{\alpha})
	/Tr(\rho_{f}\rho_{i}),
	\end{equation}
Now Eq. (10), the cyclic property, Eq. (6), and the
assumption that $\rho_{i}$ and $\rho_{f}$ commute give
	\begin{eqnarray}
	Tr(\rho_{f}\rho_{i}\tilde{C}^{\dagger}_{\alpha})
	&=&Tr(\rho_{f}\rho_{i}\Theta^{-1} C_{\alpha}\Theta)
	=Tr(\Theta\rho_{f}\Theta^{-1}\Theta\rho_{i}\Theta^{-1}
	C_{\alpha}) \nonumber \\
	&=&Tr(\rho_{f}\rho_{i}C_{\alpha})
	=Tr(\rho_{i}\rho_{f}C_{\alpha}).
	\end{eqnarray}
Therefore,
	\begin{equation}
	p(\alpha)=p(\tilde{\alpha}),
	\end{equation}
so the probabilities of CPT-related histories are equal under
the assumptions above, even without assuming that the initial
and final density matrices are the CPT reverses of each
other ($\rho_{i}=\Theta^{-1}\rho_{f}\Theta$), QED.

\baselineskip 23.6pt
As an example of a consequence of this theorem,
consider the case in which $C_{\alpha}$ is a
single string (2) with $P^{1}_{\alpha_{1}}(t_{1})$ corresponding
to low coarse-grained entropy and $P^{n}_{\alpha_{n}}(t_{n})$
corresponding to high entropy, so that the history $\alpha$ has
entropy increasing from the earliest time $t_1$ to the latest time
$t_n$.  Assuming that the definition of coarse-grained entropy
is CPT invariant, so that $\tilde{P}^{k}_{\alpha_{k}}(-t_{k})$
corresponds to the same entropy as $P^{k}_{\alpha_{k}}(t_{k})$,
then the CPT-reversed history $\tilde{\alpha}$ has entropy
decreasing from the new earliest time $-t_n$ (that of the projection
operator now adjacent to $\rho_{i}$) to the new latest time $-t_1$
(that of the projection operator now adjacent to $\rho_{f}$).
Then under the conditions above (commuting CPT-invariant
$\rho_{i}$ and $\rho_{f}$), the probability of a history $\alpha$
with one thermodynamic time asymmetry is equal to that of the
history $\tilde{\alpha}$ with the opposite thermodynamic time
asymmetry, so long as both probabilities exist.  In other words,
the asymmetry of the decoherence functional does not give any
preferred direction (in the sense of differing probabilities)
for the thermodynamic arrow of time, even if one sticks with the
convention \cite{6,7,l,8,9} that the earliest times correspond to
the operators nearest to $\rho_{i}$ in the decoherence functional.

As an aside, we may note that if $\rho_{i}$ and  $\rho_{f}$
commute so that we may construct new Hermitian
positive-semidefinite initial and final density matrices
	\begin{eqnarray}
	\rho'_{i}&=&\rho_{i}\rho_{f}/Tr(\rho_{i}\rho_{f}),\\
	\rho'_{f}&=&I,
	\end{eqnarray}
and if the set of histories $\{\alpha\}$ decoheres for the new
decoherence functional
	\begin{equation}
	D'(\alpha,\alpha')
	=Tr(\rho'_{f}C_{\alpha}\rho'_{i}C^{\dagger}_{\alpha'})
	/Tr(\rho'_{f}\rho'_{i}),
	\end{equation}
then a similar proof shows that
	\begin{equation}
	p'(\alpha)\equiv D'(\alpha,\alpha)=p(\alpha).
	\end{equation}
However, for generic commuting $\rho_{i}$ and  $\rho_{f}$,
the new decoherence condition is independent of the old one
and/or its CPT-reverse, so the theory need not be the same
for $\rho'_{i}$ and  $\rho'_{f}$ as for $\rho_{i}$ and
$\rho_{f}$ if one chooses sets of histories which decohere for
one set of density matrices and not for the other.

If we do have a final condition of indifference,
$\rho_{f}\propto I$, which corresponds to ordinary quantum
mechanics, it obviously commutes with any $\rho_{i}$ and is
CPT invariant.  Therefore, in ordinary quantum mechanics
the CPT invariance of the initial density matrix is sufficient to
imply that the probabilities of a set of histories equal the
corresponding probabilities of the CPT-reversed set (if both
sets decohere, as is necessary to get probabilities obeying
the sum rules).  Such a universe would be CPT-invariant,
according to the definition of Gell-Mann and Hartle \cite{7},
even without their alternative sufficient condition
	\begin{equation}
	\rho_{f}=\Theta\rho_{i}\Theta^{-1}.
	\end{equation}

Thus we see that the Gell-Mann--Hartle formulation of quantum
mechanics, even with greatly different commuting initial and final
conditions (such as ordinary quantum mechanics with its
final condition of complete indifference), does not by itself
give any time asymmetry for the probabilities.
It leads to CPT-symmetric universes if the initial and final
conditions are separately CPT invariant.
In this formalism, any such asymmetry in
the probabilities must lie separately within the initial and/or
final density matrix of the closed system.
This result is not in conflict with the results of Gell-Mann and
Hartle \cite{7}, who merely proposed Eq. (22) as a {\it sufficient}
condition for a CPT-invariant universe.  However,
if one regards the probabilities of decohering
sets of histories as basic and does not attach a meaning to the
entire decoherence functional (which does have an asymmetry),
one can avoid interpreting ordinary quantum mechanics
as necessarily having any time asymmetry.

Of course, the time symmetry of the probabilities of
CPT-reversed sets of decohering histories does not imply that
each history with a significant probability within one of those
sets is itself time symmetric, as was illustrated by the example
above with changing entropy.  It merely implies that the
time-reversed history in the other set has the same probability.
Thus observers in one of the histories may see that history
as being time asymmetric, even if the overall initial and final
quantum states are each separately time symmetric and so
lead to equal corresponding probabilities for the two
CPT-reversed sets of decohering histories.  This would also be
true under the alternative time-symmetric condition (22)
of Gell-Mann and Hartle \cite{7}, as they indeed carefully point
out.

Thus our observations of
an apparently time-asymmetric history for our
universe \cite{10,11} do not yet appear to rule out either
time-symmetric possibility (6) or (22), as is consistent with what
Gell-Mann and Hartle \cite{7} noted.  Possibility (6) is exemplified
by the Hartle-Hawking no-boundary proposal for the quantum
state of the universe \cite{h,hh,12,13,14}.
Emulating John A. Wheeler \cite{w}, one may say that our history
of the universe has ``time asymmetry without time asymmetry"
of the probabilities.  One can summarize the situation by
saying that not only do asymmetric boundary conditions
in the Gell-Mann--Hartle sense [inequality(7)] not necessarily imply
asymmetric probabilities, but also that symmetric conditions
[with either the Gell-Mann--Hartle equation (22) or my equation (6)]
do not necessarily imply symmetric histories.

The question now arises how to interpret the
arrow of ordinary quantum mechanics in the
formalism of Gell-Mann and Hartle.  In contrast to the
analysis above, which does not contradict any of their
results, here I shall make some speculative interpretative
comments which are my own views and are generally not
held by Gell-Mann and Hartle.  Before  doing this, I should note
that, as a consequence of the previous paragraph, both the
arrow of their formalism and the time symmetry of the probabilities
I have demonstrated (for commuting CPT-invariant initial and
final conditions) are not testable within any one individual history
of
the universe (e.g., ours) and therefore are both rather metaphysical.
Nevertheless, one can say very little if one attempts to be a
complete
positivist, and therefore I shall continue to consider how a
nonphysical
meta-observer might view the entire universe.

It seems to me that the asymmetry of the Gell-Mann--Hartle
decoherence functional has more to do with the order and
noncommutation
of the density and/or projection operators than with any
{\it time} asymmetry. It would exist even when all of these operators
are completely stationary as well as CPT invariant, in which case it
seems
very unnatural to ascribe it to anything involving time.

The asymmetry seems to get associated with time
because of the traditional rule of ordering the projection operators
in Eq. (2)
in time
order, which Gell-Mann and Hartle have adopted in their
formalism.   They do note \cite{6,7,l,8,9} that one would get an
equivalent result by a CPT tranformation of the density and
projection operators which gives them an anti-time ordering,
but they do not allow zigzags, in which the times
in the successive operators are not monotonically decreasing
or increasing.

Gell-Mann and Hartle say zigzags can lead to ``non-zero probabilities
[for] conflicting alternatives at the same time" \cite{8}.  Here
``conflicting"
cannot be taken to mean ``orthogonal,"
because then the probability sum rules would imply that the
probabilities
for conflicting alternatives would be zero, no matter what the times.
The statement is true if ``conflicting" is taken to mean
``noncommuting" \cite{p},
but then it is not clear to me why noncommuting projection operators
should be
considered as {\it not} conflicting even if they occur at different
times in
some history.

One might have thought that the probabilities for sequences of
alternatives would depend on the order in which the operators are
written down to form the string $C_{\alpha}$.  This may indeed be
true for
nonadjacent operators (or strings of them).  However, it turns out
that the order of two {\it adjacent} substrings within a string does
not affect
the probabilities (so long as they exist for both orderings), as is
shown by
the following theorem (a generalization of the penultimate sentence
of
Section III of Hartle \cite{r}):

\baselineskip 23pt
{\bf Theorem 2}:  Consider a set of histories
$\{\alpha\}=\{(\alpha_1,
\alpha_2, \alpha_3, \alpha_4)\}$ represented by
	\begin{equation}
	C_{\alpha}=c^{4}_{\alpha_{4}}c^{3}_{\alpha_{3}}
	c^{2}_{\alpha_{2}}c^{1}_{\alpha_{1}},
	\end{equation}
(where each substring $\alpha_i$ is independently allowed to take on
all
possible values)
and a corresponding zigzag set
$\{\hat{\alpha}\}=\{(\hat{\alpha}_1, \hat{\alpha}_2, \hat{\alpha}_3,
\hat{\alpha}_4)\}$
represented by
	\begin{equation}
	\hat{C}_{\hat{\alpha}}=\hat{c}^{4}_{\hat{\alpha}_{4}}
	\hat{c}^{3}_{\hat{\alpha}_{3}}
	\hat{c}^{2}_{\hat{\alpha}_{2}}\hat{c}^{1}_{\hat{\alpha}_{1}}
	=c^{4}_{\alpha_{4}}c^{2}_{\alpha_{2}}
	c^{3}_{\alpha_{3}}c^{1}_{\alpha_{1}}
	\end{equation}
with $c^{2}_{\alpha_{2}}$ and $c^{3}_{\alpha_{3}}$ interchanged.
Then if both sets decohere, the corresponding probabilities are
equal,
	\begin{equation}
	p(\tilde{\alpha})=D(\tilde{\alpha},\tilde{\alpha})
	=p(\alpha)=D(\alpha,\alpha).
	\end{equation}

{\bf Proof}:  To abbreviate the notation, let
	\begin{equation}
	c_i = c^i_{\alpha_{i}},\;\;
	c'_i = \sum_{\alpha'_i\neq\alpha_i} c^i_{\alpha'_{i}}
	= I - c_i.
	\end{equation}
Then the weak decoherence condition (4) implies
	\begin{eqnarray}
	0&=&Re\,Tr(\rho_f c_4 c_3 c_2 c_1 \rho_i
	c_1^{\dagger} c_2^{\dagger} c_3^{\prime\dagger}
c_4^{\dagger})\nonumber\\
	&=&Re\,Tr(\rho_f c_4 c_3 c_2 c_1 \rho_i c_1^{\dagger}
c_2^{\dagger}
	c_4^{\dagger})
	-p(\alpha) ,\\
	0&=&Re\,Tr(\rho_f c_4 c_3 c'_2 c_1 \rho_i
	c_1^{\dagger} c_2^{\dagger} c_3^{\dagger}
c_4^{\dagger})\nonumber\\
	&=&Re\,Tr(\rho_f c_4 c_3 c_1 \rho_i c_1^{\dagger}
	c_2^{\dagger} c_3^{\dagger} c_4^{\dagger})-p(\alpha) ,\\
	0&=&Re\,Tr(\rho_f c_4 c_3 c'_2 c_1 \rho_i
	c_1^{\dagger} c_2^{\dagger} c_3^{\prime\dagger}
c_4^{\dagger})\nonumber\\
	&=&Re\,Tr(\rho_f c_4 c_3 c_1 \rho_i c_1^{\dagger}
	c_2^{\dagger} c_4^{\dagger})+\nonumber\\
	&&- Re\,Tr(\rho_f c_4 c_3 c_2 c_1 \rho_i
	c_1^{\dagger} c_2^{\dagger} c_4^{\dagger})+\nonumber\\
	&&- Re\,Tr(\rho_f c_4 c_3 c_1 \rho_i
	c_1^{\dagger} c_2^{\dagger} c_3^{\dagger}
c_4^{\dagger})+p(\alpha).
	\end{eqnarray}
Combining Eqs. (27)-(29) gives
	\begin{equation}
	p(\alpha)=Re\,Tr(\rho_f c_4 c_3 c_1 \rho_i c_1^{\dagger}
c_2^{\dagger}
c_4^{\dagger}).
	\end{equation}
Similarly, the corresponding weak decoherence condition
$Re\,D(\hat{\alpha},
\hat{\alpha}')=0$
for $\hat{\alpha}\neq\hat{\alpha}'$ gives
	\begin{equation}
	p(\hat{\alpha})=Re\,Tr(\rho_f c_4 c_2 c_1 \rho_i
c_1^{\dagger} c_3^{\dagger}
c_4^{\dagger}).
	\end{equation}
Now the cyclic property of the trace allows us to move $\rho_f$ to
the right
end of the matrix of Eq. (31), and then this matrix is the Hermitian
conjugate
of the matrix of Eq. (30).  Thus the real parts of the traces are
equal, Eq.
(25), QED.

\baselineskip 23pt
Note that the $c_i$'s can be projection operators, or strings of
them, or even
sums of strings, but we do need a coarse graining of $\{\alpha\}$ to
include
the four histories represented by $C_{\alpha}=c_4 c_3 c_2 c_1$,
$c_4 c'_3 c_2 c_1$, $c_4 c_3 c'_2 c_1$, and $c_4 c'_3 c'_2 c_1$
(not just the two histories $C_{\alpha}$ and $I - C_{\alpha}$), and
similarly
for $\{\hat{\alpha}\}$.

To interchange two nonadjacent substrings or sums of strings and get
the same
probabilities, we would need three permutations to get them through
the
intermediate substring and through each other.  Without assuming
that the two intermediate permutations also give decohering sets of
histories,
the decoherence of merely the initial and final sets is in many cases
sufficient
for proving the equality of their corresponding probabilities, but
not always
\cite{p}.
Thus a difference in the probabilities appears to be possible.

Therefore, except possibly for the caveat of the last paragraph, the
motivation
to exclude zigzags and keep the projection operators in time (or
anti-time)
order is lost on me.  Thus I am not convinced that the asymmetry that
arises
from the order of the projection operators relative to that of
the density matrices should be
associated with the order of time.  In other words, I do not see that
ordinary
quantum mechanics with CPT-invariant initial conditions
gives any {\it time} asymmetry, at least for the probabilities of an
CPT-reversed pair of decohering sets of histories,
although in a different sense one could say it is indeed
quantum mechanics that allows nonunique histories, each of which can
be time
asymmetric even when the whole set of CPT-reversed pairs is not.

{\bf Acknowledgments}:  This paper was motivated by
discussions with Murray Gell-Mann and James Hartle
and was phrased more precisely as a result of many further
discussions
with them, for which I am deeply grateful,
but it by no means represents their interpretation or meaning
of time asymmetry, despite the fact that there is no direct
contradiction between our basic results.
A comment by Stephen Hawking on how his
superscattering matrix for black hole formation and
evaporation can lead to loss of coherence without being time
asymmetric was remembered after the first part of this work
was conceived and may have had a subconscious influence.

The hospitality of the University of California at Santa
Barbara, of Northwest Airlines over the Western United
States, and of Nelson and Zena Page in Liberty, Missouri,
USA, where this work was first formulated and written up, is
gratefully acknowledged.  Financial support was provided in
part by the Natural Sciences and Engineering Research
Council of Canada.

\end{document}